# Immense Fidelity Enhancement of Encoded Quantum Bell Pairs at Short and Long-distance Communication along with Generalized Design of Circuit

Syed Emad Uddin Shubha[1], Md. Saifur Rahman[2], M.R.C. Mahdy[1*]

[1] *Department of Electrical & Computer Engineering, North South University, Bashundhara, Dhaka 1229, Bangladesh*

[2] *Department of Electrical & Electronic Engineering, Bangladesh University of Engineering and Technology, Dhaka, Bangladesh*

Corresponding author's email address: [*]mahdy.chowdhury@northsouth.edu

**Quantum entanglement is a unique criterion of the quantum realm and an essential tool to secure quantum communication. Ensuring high-fidelity entanglement has always been a challenging task owing to interaction with the hostile channel environment created due to quantum noise and decoherence. Though several methods have been proposed, achieving almost 100% error correction is still a gigantic task. As one of the main contributions of this work, a new model for 'large distance communication' has been introduced, which can correct all bit flip errors or other errors quite extensively if proper encoding is used. To achieve this purpose, at the very first step, the idea of differentiating the 'long-distance communication' and 'short-distance applications' has been introduced. Short-distance is determined by the maximum range of applying unitary control gates by the qubit technology. As far as we know, there is no previous work that distinguishes long and short distance applications. At the beginning, we have applied stabilizer formalism and Repetition Code for decoding to distinguish the error correcting ability in long and short distance communication. Particularly for short-distance communication, it has been demonstrated that a 'properly encoded' bell state can identify all the bit flip, or phase flip errors with 100% accuracy theoretically. In contrast, if the bell states are used in long-distance communication, the error-detecting and correcting ability reduces at huge amounts. To increase the fidelity significantly and correct the errors quite extensively for long-distance communication, a new model based on classical communication protocol has been proposed. According to our proposal, for both short and long-distance communication,** *proper encoding method* **should be implemented. All the required circuits in these processes have been generalized for arbitrary (even) numbers of ancilla qubits during encoding. Proposed analytical results have also been verified with the Simulation results of IBM QISKIT QASM.**

## 1. Introduction

Quantum entanglement is a nonlocal property of quantum mechanics (extensively investigated by J. Bell [1]), which was initially pointed out to oppose quantum mechanics in a famous EPR paper by Einstein et al. [2]) Quantum Entanglement is a crucial part of Quantum Communication now a days. For example, Quantum teleportation [3], superdense coding [4], Ekert Protocol [5] for Quantum Key Distribution, etc., rely on establishing entanglement pair between two nodes in a quantum communication network.

Ensuring high fidelity quantum entanglement between two nodes separated by longer distance has always been challenging. To overcome the challenge, first generation Quantum Repeater has been proposed that utilizes entanglement distribution and swapping protocol along with entanglement purification method [6]. In order to understand the mechanism in details, first, entanglement links are created between network nodes (Entanglement Distribution). After that, a bell basis measurement is performed within a node on two qubits which are halves of separate bell states, allowing to provide a longer entanglement link connecting adjacent repeater nodes (Entanglement Swapping). Finally, more highly entangled state is created from a number of lower quality ones (Entanglement Purification) [7-11].

In previous works, entanglement purification has been proposed to increase the fidelity in long distance. But to begin, the communication rate decreases polynomially with distances, thus becoming very slow for long-distance classical communication. Furthermore, multiple entangled pairs are required and it takes many rounds of purification to obtain high fidelity bell pairs. Most importantly, the success rate is very low. If it is failed, then we need to start again. [12]

Now let's assume Alice and Bob share two copies of $|\phi_+\rangle$ state. In entanglement purification protocol, at the beginning, Alice and Bob apply CNOT gate in their side of qubits. After that they measure one qubit in each side and then compare the results using classical communication [7]. If the results match; then the fidelity of unmatched bell pair is increased, else the fidelity decreases, and they need to throw away the qubits and restart the whole process [13-15]. In this case, there is a 50% probability of fidelity being decreased, which is a major drawback.

At the very first step of this work, the idea of differentiating the 'long-distance communication' and 'short-distance applications' has been introduced. The definition of long and short distance communication depends on the qubit technology used in the application. If the laboratory supports applying controlled unitary gate at distance x, then anything smaller than that is distinguished as short-distance and anything larger than that is considered as long-distance communication. The main reason behind this work is that entanglements can be used in laboratory applications as well. How the same channel impacts the Bell states and how the Error Correcting abilities changes with distance require proper studies. As far as we know, there was no work that distinguishes long and short distance communication. To the best of our knowledge, there is no available report in literature with the 100% success rate of error correction.

To solve the error issues of all the aforementioned cases (i.e., channels), we have not only applied the stabilizer formalism and quantum repetition code in our studied cases but also investigated the physical meaning. We have provided a new model as well. We have also simulated them in QISKIT QASM Simulator.

However, in this work, we have shown that Quantum Error Correcting Codes (QECCs) instead of entanglement purification improves the fidelity of Bell pairs significantly. Using stabilizer formalism, and proper encoding for channel, we have shown immense increase in fidelity, where we can correct all the bit flip or phase flip errors particularly for short distance. We have studied both the short and long-distance communication application separately, which is not explored before us to the best of our knowledge.

In this work, we have encoded Bell States using Quantum Repetition Code and calculated the fidelity in bit flip and phase flip channels separately. Note that if a QECC can correct both bit flip and phase flip errors, then it can correct any arbitrary single qubit errors [16].

For arbitrary channel, we can use nested encoding method to correct errors. We have examined two cases for entangled state in this paper. In the first case, the entangled state is local, and in the second case they are separated by a long-distance to make communication. There are two subcases in the second case, with/ without classical communication channel.

We have shown that for local entanglement states (in short-distance communication), it is possible to distinguish and correct all possible bit flip or phase flip errors. For long-distance communication, some fidelities are lost due to the requirement of non-local gates.

A novel model has been proposed for long-distance communication. It has also been shown by us that using classical communication and local measurement, it is possible to correct all the bit flip and phase flip errors in long-distance communication.

A generalized algorithm for building a $(2k + 1,1)$ repetition code decoder circuit is discussed. We have shown that if we want to use entanglement as a resource for long-distance communication, then fidelity is limited by the repetition code-based scheme's fidelity, and entanglement states in general are not too different from product states. The advantage due to entanglement comes to light only if an entangled system is local.

In the upcoming section, at first, we will go through some preliminary concepts to understand our work properly. In next two sections, we have first reviewed our generalized results and circuits for QRC based encoding-decoding, which is applicable for both short and long-distance communication, followed by the decoding process using stabilizer formalism, which contains protocol for short-distance communication, and our proposed novel scheme for long distance communication. At the end, we discussed how the mutual information in qubits can help in error correction.

We have compiled the workflow of our paper in Figure 1 and Figure 2 displays our explored configurations of bell states. Figure 3 contains our proposed protocol for long-distance communication with classical channel.

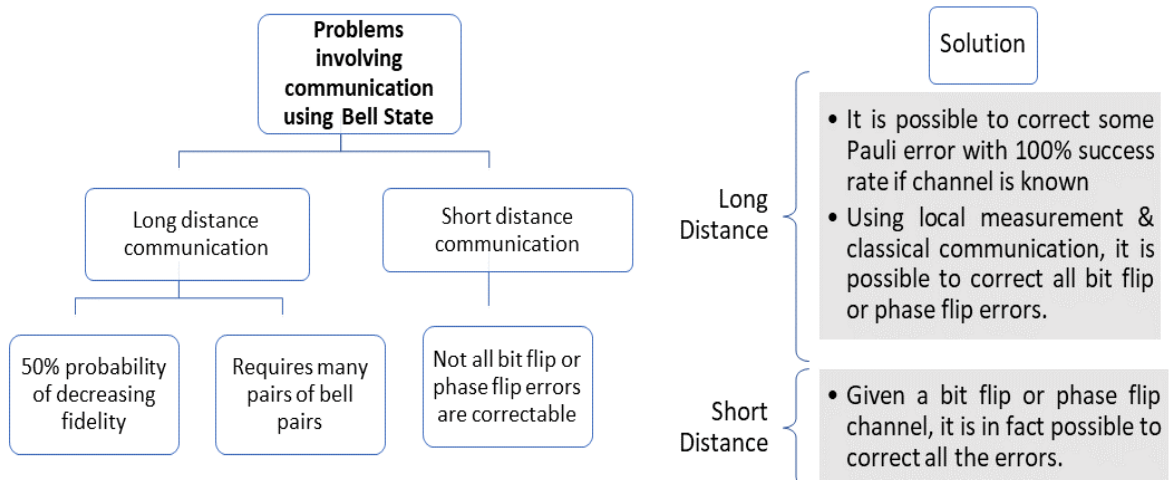

Fig. 1: Flowchart of our work. Generalized circuit designs are provided in order to implement proposed schemes.

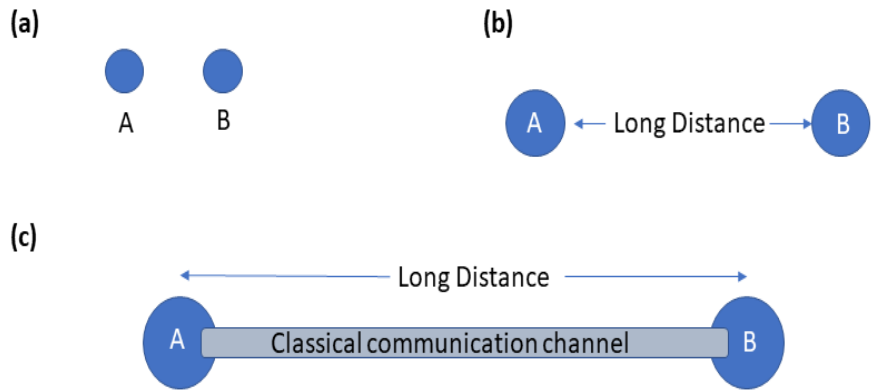

Fig. 2: A and B are two spin ½ entangled particles. Three cases are studied in this work. (a) Local (Short-distance) Entangled State (b) Entangled State separated by long-distance (c) Entangled State separated by long-distance but with a classical communication channel. The short distance means the laboratory capacity to perform a distant controlled unitary operation.

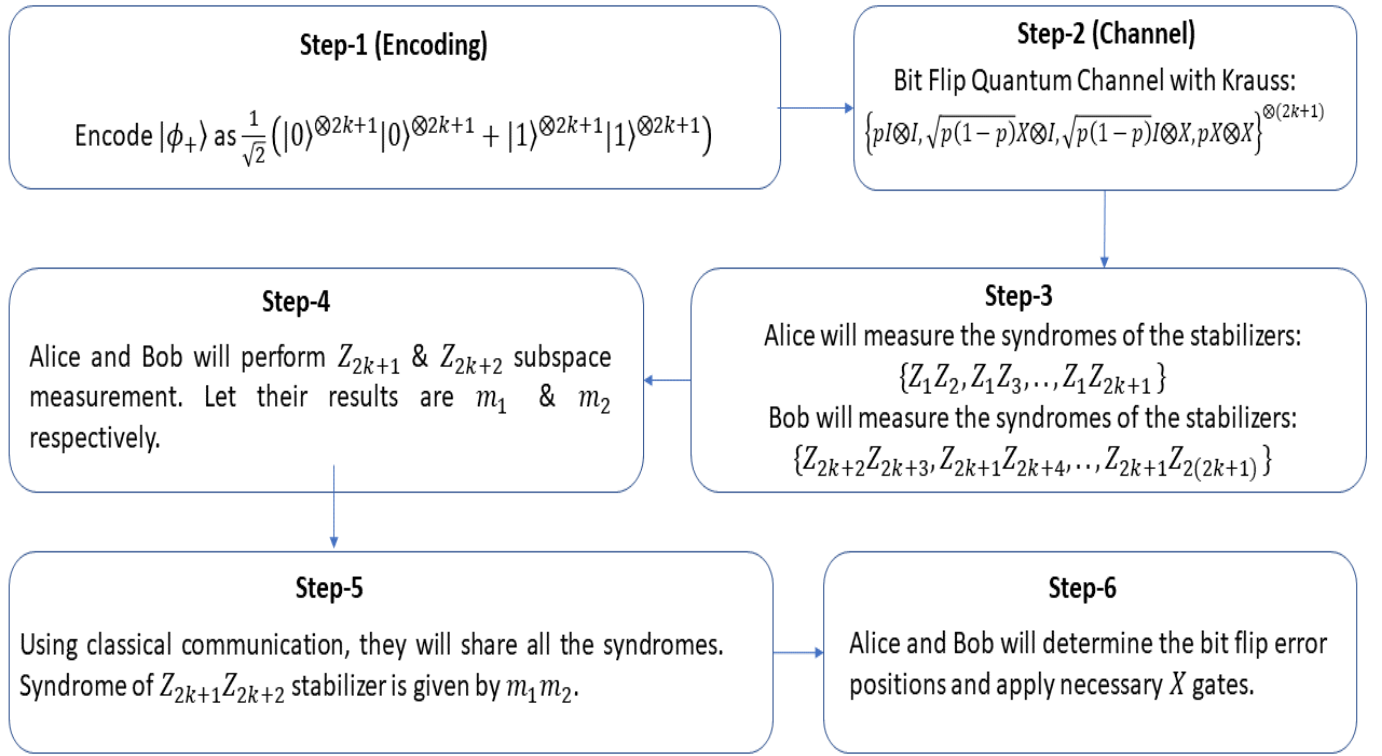

Fig. 3: Proposed protocol for correcting all bit flip or phase flip error in long-distance communication. Classical communication and local measurement are used to determine all the syndromes for error correction.

## 2. Preliminary Concepts

### 2.1 Quantum States, Density Matrices, Quantum Logic Gates, and Bell States

Qubits are the fundamental building blocks of Quantum Computing. A qubit represents the state of a two-level quantum system. The general one qubit state, $|\psi\rangle \in \mathcal{H}$ can be written as ($0 \leq \theta \leq \pi, 0 \leq \phi \leq 2\pi$, $\mathcal{H}$ is associate Hilbert space): [16]

$$|\psi\rangle = \cos\left(\frac{\theta}{2}\right)|0\rangle + e^{i\phi}\sin\left(\frac{\theta}{2}\right)|1\rangle \tag{1}$$

The set $\{|0\rangle, |1\rangle\}$ and $\{|+\rangle, |-\rangle\}$ represent Z basis (computational basis) and X basis (diagonal basis) respectively, are related by a unitary operator, known as Hadamard matrix, $H$ as:

$$H|0\rangle = |+\rangle, \ H|1\rangle = |-\rangle \tag{2}$$

In Z basis, the Hadamard matrix can be written as:

$$H = \frac{1}{\sqrt{2}}\begin{pmatrix} 1 & 1 \\ 1 & -1 \end{pmatrix} \tag{3}$$

Density matrices, $\rho \in \mathcal{B}(\mathcal{H})$ are the general form of Quantum states, which are positive semi-definite matrices ($\rho \geq 0$) with the normalization condition, $Tr[\rho] = 1$.

Quantum logic gates are unitary transformations If $U$ is a unitary operator then $U^\dagger = U^{-1}$ and it acts on state $|\psi\rangle$, then the final state is given by $|\psi'\rangle = U|\psi\rangle$. For density matrix, we can express the operation as:

$$\rho' = U\rho U^\dagger \tag{4}$$

Some important single qubit gates are given below [13,16]:

(i) *The bit flip gate (Pauli X gate)*: This is quantum analog of classical NOT gate, in computational basis, i.e.,

$$X|0\rangle = |1\rangle, \ X|1\rangle = |0\rangle \tag{5}$$

(ii) *The phase flip gate (Pauli Z gate)*: This unitary gate adds a 180° phase if acts on $|1\rangle$ state, and doesn't change the $|0\rangle$ state, i.e.,

$$Z|0\rangle = |0\rangle, Z|1\rangle = -|1\rangle \tag{6}$$

Z gate works as a bit-flip gate on a diagonal basis.

The Hadamard gate, Pauli X and Z gates are related by,

$$HZX = X \tag{7}$$

The most important class of unitary gate for multipartite quantum states is Controlled Unitary gate. Let's assume we have $(n + 1)$ qubits named as $0, 1, 2, \ldots, n$. A single qubit unitary operation $U$ is applied on $n^{th}$ qubit when first $n$ qubit is in the basis state $|1\rangle^{\otimes n}$. We can express the operation of the $C^n U$ gate as:

$$C^n U = I^{\otimes n+1} + |1\rangle^{\otimes n}\langle 1|^{\otimes n} \otimes (U - I) \tag{8}$$

For example, Controlled NOT gate is defined as,

$$CNOT|x\rangle|y\rangle = |x\rangle|x \oplus y\rangle, \quad x, y \in \{0,1\} \tag{9}$$

Here, $|x\rangle$ is called control qubit and $|y\rangle$ is called target qubit. If $U$ is Pauli $X$ gate, then $C^n X$ gate is called Toffoli gate in general.

Now, we shall look at entanglement and Bell state. System A, B are called entangled state if their combined density matrix $\rho_{AB}$ cannot be written as $\rho_A \otimes \rho_B$. Bell States are a special type of entangled state for bipartite quantum system.

Bell states are defined as,

$$|\phi_\pm\rangle = \frac{1}{\sqrt{2}}(|00\rangle \pm |11\rangle), |\psi_\pm\rangle = \frac{1}{\sqrt{2}}(|01\rangle \pm |10\rangle)$$

We can combine these four states as [17],

$$|\psi_{mn}\rangle_{AB} = \frac{1}{\sqrt{2}}(|0\rangle_A|m\rangle_B + (-1)^n|1\rangle_A|1 \oplus m\rangle_B) \tag{10}$$

Where $m, n \in \{0,1\}$ and $|\phi_+\rangle = |\psi_{00}\rangle, |\phi_-\rangle = |\psi_{01}\rangle, |\psi_+\rangle = |\psi_{10}\rangle, |\psi_-\rangle = |\psi_{11}\rangle$

Bell states are called maximally entangled state since the entanglement entropy is maximum. Fig. 4 shows how to create $|\phi_+\rangle$ state using Hadamard and CNOT gate:

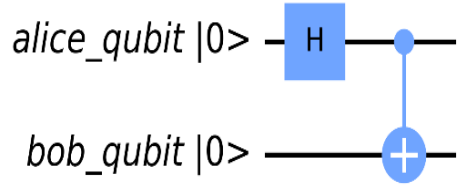

Fig. 4: Creating Bell state. Hadamard gate is applied to keep the control qubit in superposition state.

We can easily convert one bell state to another by performing some unitary operation:

$$(I \otimes Z)|\phi_+\rangle = |\phi_-\rangle, (I \otimes X)|\phi_+\rangle = |\psi_+\rangle, (I \otimes XZ)|\phi_+\rangle = |\psi_-\rangle \tag{11}$$

Transformations in Eq. (11) play a major role in Superdense Coding [4].

## 2.2 Quantum Channel, CPTP Maps, Fidelity

An open quantum system interacts with the environment and gets entangled very easily. Let's imagine we have prepared a quantum state $\rho_S = |\psi\rangle\langle\psi|$ and the environment (or channel) has the quantum state $\rho_E = |\varepsilon\rangle\langle\varepsilon|$. Initially, the overall quantum state is given by: $\rho_{SE} = \rho_S \otimes \rho_E$. Now interaction with the channel will

induce some unitary evolution $U_{SE}$ in total state $\rho_{SE}$. After the evolution, the state is given by, $\rho'_{SE} = U_{SE}\rho_{SE}U_{SE}^\dagger = U_{SE}\rho_S \otimes \rho_E U_{SE}^\dagger$ [12-13].

Now if we want to deal with our prepared system, we need to trace out the environment part. Thus, the evaluated system will become:

$$\rho'_E = Tr_E[\rho'_{SE}] = \sum_k E_k \rho_S E_k^\dagger \tag{12}$$

$\rho'_E$ cannot be written as product state in general, hence we call that quantum state is entangled with environment. Here $E_k = \langle e_k | U_{SE} | \varepsilon \rangle$. We note that, $\sum_k E_k^\dagger E_k = 1$, $\rho'_E \geq 0$ with $Tr[\rho'_E] = 1$. Therefore, the evolution in Eq. (12) results in a valid density matrix fulfilling all the properties. Hence, we call this operation a Completely Positive Trace Preserving (CPTP) map (also known as Super operators). We call $\{E_k\}$ the Krauss Operator of CPTP Map. [25]

Eq. (12) gives the most general quantum evolution. For example, for unitary evolution, the set of Krauss operator is given by $\{U\}$, where $U$ is a unitary gate. Measurement, tracing out a subsystem also can be represented using CPTP Maps.

Now, let's unitary $U_i$ is applied on $\rho$ with probability $p_i$. Then, we deduce the new density matrix, $\rho' = \sum_i p_i U_i \rho U_i^\dagger$, where the set of Krauss operators $\{E_i\}$, is given by:

$$E_i = \sqrt{p_i} U_i \tag{13}$$

We can define bit flip channel for single qubit as: $\rho = p.X|\psi\rangle\langle\psi|X + (1-p)|\psi\rangle\langle\psi|$, thus Eq. (13) gives the Krauss operators, $E_{bitflip} = \{\sqrt{p}X, \sqrt{1-p}I\}$. Similarly, for phase flip channel, the Krauss operators are, $E_{phaseflip} = \{\sqrt{p}Z, \sqrt{1-p}I\}$.

Finally, we will define fidelity. Suppose we have prepared a pure quantum state $|\psi\rangle\langle\psi|$, which becomes a mixed state $\rho$ due to interaction with nature according to Eq. (15). We then define the fidelity as:

$$F = \sqrt{\langle\psi|\rho|\psi\rangle} \tag{14}$$

We note that, $0 \leq F \leq 1$ and the equity holds only if the state remains pure.

## 2.3 Concept of Pauli Group and Stabilizer

The set $\wp^n = \{+1, -1, +i, -i\} \times \{I, X, Y, Z\}^{\otimes n}$ forms a Group under matrix multiplication, known as Pauli Group. Here, $X = \sigma_1 = \begin{pmatrix} 0 & 1 \\ 1 & 0 \end{pmatrix}$, $Y = \sigma_2 = \begin{pmatrix} 0 & -i \\ i & 0 \end{pmatrix}$, $Z = \sigma_3 = \begin{pmatrix} 1 & 0 \\ 0 & -1 \end{pmatrix}$ are called Pauli Matrices. We also denote $I = \sigma_0$ as convention, to include the multiplicative identity [18, 23].

A stabilizer group $S \leq \wp^n$ is a Pauli subgroup with some properties.

Let's $p_1, p_2, p \in S$, then (i) $p^2 = I^{\otimes n}$ (ii) $[p_1, p_2] = 0$ (iii) $-I \notin S$.

We often use generators to describe a group. For example, $S = \{\mathbb{I}, Z_1Z_2, Z_2Z_3, Z_3Z_1\}$ is a stabilizer generated by $Z_1Z_2, Z_2Z_3$. We write $S = \langle Z_1Z_2, Z_2Z_3 \rangle$.

Here the notation $M_r = I^{\otimes(r-1)} \otimes M \otimes I^{\otimes(n-r)}$ and $\mathbb{I}$ means $I^{\otimes n}$ for n-fold Pauli group.

In stabilizer formalism, we first denote the coding space, $\mathcal{H}_c$, using stabilizer set, say $S$. Pauli error (elements of Pauli Group) either will commute with elements of $S$ or anti-commute. Now if the generator set of $S$ has cardinality $n$, then we can quickly identify errors by checking whether it commutes with each element of $S$. We then call it Syndrome.

Let's $E$ is a Pauli Error and $S = \langle s_1, s_2, \ldots, s_n \rangle$ is a stabilizer of $\in \mathcal{H}_c$. Then,

$$s_i |\psi\rangle = |\psi\rangle \; \forall |\psi\rangle \in \mathcal{H}_c \tag{15}$$

Let's $s_i E = k_i E s_i$, where $k_i \in \{-1, 1\}$. If $k_i = 1$, we say $E$ commute with $s_i$, and if $k_i = -1$, we say $E$ anti-commute with $s_i$. The set $\{k_i\}$ is called the syndrome of $E$.

Now we note that, $s_i(E|\psi\rangle) = k_i E s_i |\psi\rangle = k_i E |\psi\rangle$, therefore $E|\psi\rangle$ is in the $k_i$ eigenspace of $s_i$. There are $2^n$ possible errors we can distinguish and each of these errors divide the total Hilbert space into equidimensional subspaces.

The subcircuit to measure the syndrome is given in Figure 5 [22]:

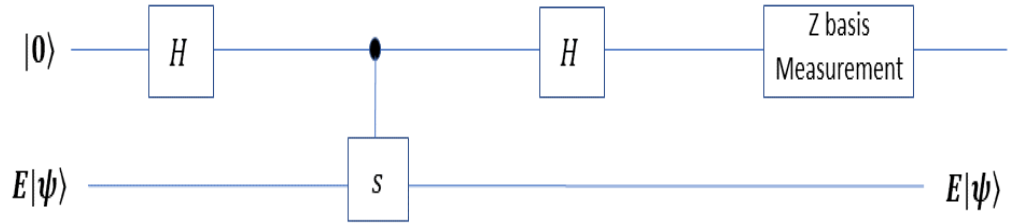

Fig. 5: Subspace Measurement in stabilizer formalism. By measuring ancilla qubit, we can decide whether it is in +1 or -1 eigenspace of stabilizer s.

The operation of the circuit is explained below:

1) First, we apply $I \otimes H$ in state $|0\rangle E|\psi\rangle$, thus the resultant state is, $\frac{1}{\sqrt{2}}(|0\rangle E|\psi\rangle + |1\rangle E|\psi\rangle)$
2) Now we apply controlled stabilizer gate, $U = |0\rangle\langle 0| \otimes I + |1\rangle\langle 1| \otimes s$, the resultant state is,
$$\frac{1}{\sqrt{2}}(|0\rangle E|\psi\rangle + |1\rangle s E|\psi\rangle) = \frac{1}{\sqrt{2}}(|0\rangle E|\psi\rangle + k|1\rangle E|\psi\rangle)$$
3) Then we again apply $I \otimes H$ and then measure the ancilla qubit. The result is given by,
$$\frac{k+1}{2}|0\rangle E|\psi\rangle + \frac{k-1}{2}|1\rangle E|\psi\rangle$$

We note that if $k = 1$ then the total state will collapse into $|0\rangle E|\psi\rangle$ and if $k = -1$ then the collapsed state is $|1\rangle E|\psi\rangle$. Thus, by measuring ancilla qubit, we can determine the relevant syndrome with respect to the stabilizer.

## 3. Bipartite Quantum States in Bit Flip and Phase Flip Channel (For Both Long and Short-distance Communication)

In this section we will briefly state our generalized circuit design and results for QRCs. We will encode as well as decode Bell pairs using QRCs. This result is applicable for both short and long-distance communication. In the next section, we will encode using QRC but decode using stabilizer syndromes, which will be applicable for short distances and our newly proposed scheme for long-distance communication with classical communication channel.

### 3.1 Without Encoding

To understand the effect of Pauli Errors on Bipartite Quantum States, we will first calculate the fidelity without any encoding. In section 2.2, we have described how to model bit flip channel for single qubit. For bipartite quantum state, we have modelled the bit flip channel as: $E^{bf} = \{\sqrt{p}X, \sqrt{1-p}I\}^{\otimes 2} = \{(1-p)I \otimes I, \sqrt{p(1-p)}X \otimes I, \sqrt{p(1-p)}I \otimes X, pX \otimes X\}$, where $p$ denotes the probability of single bit flip.

Considering arbitrary bipartite state and a bell state we can derive the following equations for minimum fidelity ($F_1$ and $F_2$ respectively):

$$F_1 = p, F_2 = \sqrt{p^2 + (1-p)^2} > F_1 \tag{16}$$

Hence, we note that maximally entangled states are less prone to errors. We will, however, try to improve the fidelity using Quantum Error Correction in the next section.

For phase flip channel, we can easily convert phase flip errors into bit flip errors using Eq. (7).

### 3.2 Quantum Repetition Code and General Encoding-Decoding Process for Single Qubit

The repetition code, which is an error correction code, is used in the classical communication channel. Later, it was used for 3-qubit codes in quantum communication by A. Peres [15]. In general, for $(2k+1,1)$ repetition code, we encode $|0\rangle$ as $|0\rangle^{\otimes(2k+1)}$ and $|1\rangle$ as $|1\rangle^{\otimes(2k+1)}$. For this, we need $2k$ ancilla qubits initialized at $|0\rangle$ state. Then we use controlled NOT gate by keeping the main qubit in control and ancilla qubits as target. Note that, a $(2k+1,1)$ repetition code can correct $k$ bit flip errors. For phase flip errors, we encode for bit flip channel then use figure 5 to convert phase flip errors into bit flip errors.

Figure 6 shows the encoding procedure in 3 qubit repetition code in terms of quantum circuit:

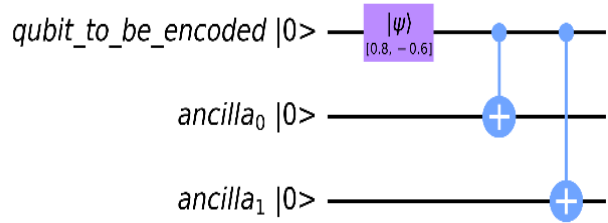

Fig. 6: Encoding in Quantum Repetition Code (3 qubit repetition code). After encoding, the state $(0.8|0\rangle - 0.6|1\rangle)$ is encoded as $(0.8|000\rangle - 0.6|111\rangle)$.

If a channel is good (probability of single qubit bit flip is $p < 0.5$), then majority of errors are in correctable range [22, 24, 25].

In this section we will provide a general scheme for designing any QRC circuit. This scheme will be used to derive generalized result of fidelity for both single qubit and bipartite quantum states.

The general scheme is given below:

(i) **Encoding:** Take $2k$ ancilla qubits (numbered as $1,2,..,2k$) in state each initialized at $|0\rangle$. At this stage, we can denote the state as $|\psi\rangle\langle\psi| \otimes |0\rangle^{\otimes 2k}\langle 0|^{\otimes 2k}$.
Now apply CNOT gates in these ancilla qubits keeping the $0^{th}$ qubit (with state $|\psi\rangle$) in control. The operation of unitary operator $U_{enc}$ is given by Eq. (17) [$q_k \in \{0,1\} \forall k$]:

$$U_{enc} = \sum_{q_0,q_1,...,q_{2k}} \left[ \left( \bigotimes_{j=1}^{2k} |q_j \oplus q_0\rangle\langle q_j| \right) |q_0\rangle\langle q_0| \right] \quad (17)$$

(ii) **Channel:** Now the encoded qubits are passed in a bit flip channel. The Krauss operators are modelled by us as,

$$E^R = \{\sqrt{p}X, \sqrt{1-p}I\}^{\otimes 2k+1} \quad (18)$$

If the channel is a phase flip channel, then we can apply Hadamard gate before and after passing through the channel.

(iii) **Decoding:** Decoding can be done in two steps.

(a) **Step 1:** Apply $U_{enc}^\dagger$. Now note that if no error is occurred in the $0^{th}$ qubit, then the state is $0^{th}$ qubit is already error free. But if there was an error then there is no way to distinguish the effect. But we know $(2k + 1, 1)$ repetition code can correct $k$ bit flip errors.

(b) **Step 2:** We note that after step 1, for most probable errors (less than $k$ errors in these $2k + 1$ qubits) if there is an error in the $0^{th}$ qubit, then at more than $k$ ancilla qubits will be at state $|1\rangle$. The required gate is given by:

$$U_{last} = X \otimes \sum_{j>k} \left( E_{i_jj}|0\rangle^{\otimes 2k} \right)\left( E_{i_jj}|0\rangle^{\otimes 2k} \right)^\dagger \quad (19)$$

Here $E_{i_jj}$ is bit flip error in $j$ ancilla qubits. For example, for 5 qubit repetition code, $k = 2$. So $j = 3,4$. For j=3, $E_{i_jj}$'s are given by $\{X \otimes X \otimes X \otimes I, X \otimes X \otimes I \otimes X, X \otimes I \otimes X \otimes X, I \otimes X \otimes X \otimes X\}$, and for j=4, $E_{i_jj}$'s are given by $\{X^{\otimes 4}\}$. There is total $\binom{2k}{j}$ of such $E_{i_jj}$.

Another way is measuring all the ancilla qubits in $Z$ basis after step 1 and if more than $k$ ancilla qubits are measured as $-1$, then all then we can simply add an $X$ gate in the $0^{th}$ qubit.

Fig. 7 provides an example for decoder circuit (five qubit repetition code).

Using Eq. (12) we find the resultant state, then using described decoding technique, tracing out ancilla qubits, we obtain the following result:

$$\rho' = P|\psi\rangle\langle\psi| + (1-P)X|\psi\rangle\langle\psi|X \quad (20)$$

P is given by Eq. (21). Therefore, we obtain the expression for minimum fidelity in Eq. (22) for encoding single qubit with $(2k+1)$ repetition code:

$$P = \sum_{r=0}^{k} \binom{2k+1}{r} p^r (1-p)^{2k+1-r} \quad (21)$$

$$F_{r,min} = \sqrt{\sum_{r=0}^{k} \binom{2k+1}{r} p^r (1-p)^{2k+1-r}} \quad (22)$$

On the other hand, without encoding, the fidelity becomes $F_{nr,min} = \sqrt{p} < F_{r,min}$ for $p < 0.5$ (good channel), thus giving application of repetition code more usefulness.

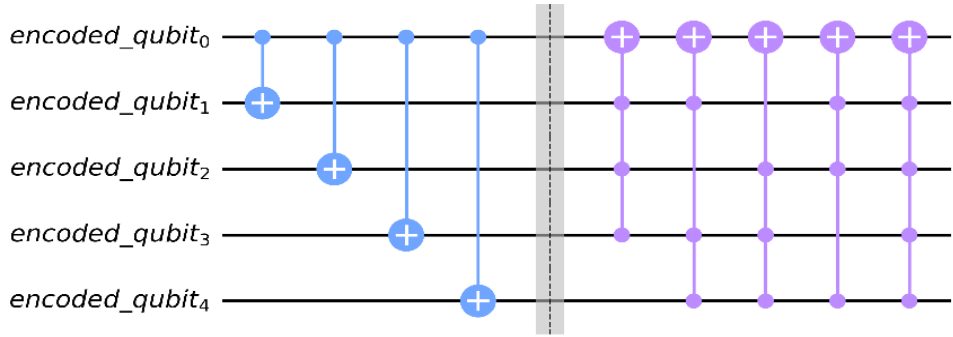

Fig. 7: Decoder circuit for 5 qubit repetition code. Here blue colored gates combined results in $U_{enc}^\dagger$ and the violate colored gates function as $U_{last}$ as a whole.

## 3.3 General Results for Applying Quantum Repetition Code in Bipartite State (For Both Long and Short-distance Communication)

Now we shall discuss how to implement QRC Error Correction for bipartite state. Let's consider two party, Alice and Bob with spin ½ particle A and B respectively. They can be separated by any distance (long or short). The quantum state of the composite system AB can be a bell state before encoding. We can use the encoding and decoding subcircuit of Fig. (5), (6) to establish a repetition code-based Quantum Error Correction for bipartite quantum state. Both Alice and Bob will take separate ancilla qubits, each initialized at state $|0\rangle$, the number of ancilla qubits are $2k$ for each. Now they will encode their qubits separately following the scheme described in section 3.2. After encoding, the channel (assume bit flip channel) will introduce some quantum noise, and some of the noise can be corrected by applying decoding operation (also described in 3.2) separately. The whole process is illustrated in Fig. 8.

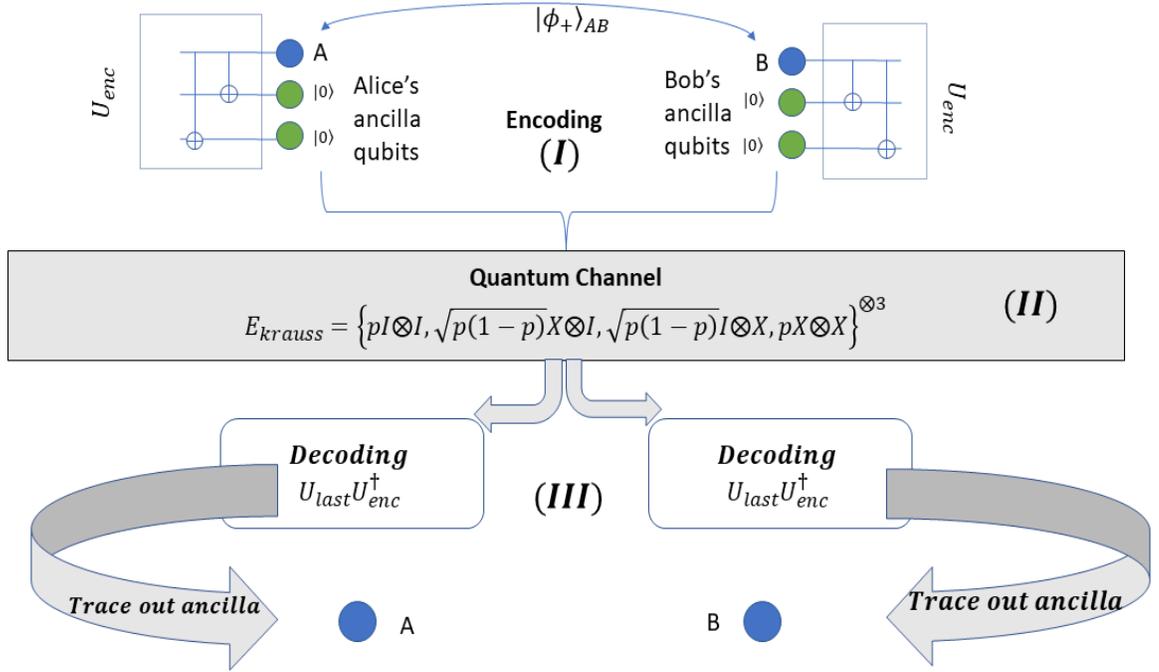

Fig. 8: Alice and Bob both share a bell state. They need to encode their qubit by taking ancilla qubits and applying $U_{enc}$ both side (I). After that they can send them in a quantum bit flip channel (II). After the operation of channel, both Alice and Bob will apply $U_{last}U_{enc}^{\dagger}$ to decode their qubits separately (III). The Fidelity lost due to channel interaction in (II) will increase again in (III) due to error correction operation.

For example, we can encode $|\phi_+\rangle$ state as, $\frac{1}{\sqrt{2}}\left(|0\rangle^{\otimes 2k+1}|0\rangle^{\otimes 2k+1} + |1\rangle^{\otimes 2k+1}|1\rangle^{\otimes 2k+1}\right)$. Now we will discuss our simulation procedure in brief.

Firstly, we created a bell state and then encoded the state using repetition code. After that it was passed through a bit-flip channel (modelled by X gates between barrier) and then decoded as described earlier. Finally, we have made a bell basis measurement in order to calculate the fidelity. A probabilistic model was used based on Monte Carlo method to perform the CPTP Map, given by the Krauss Operator: $E^b = \{\sqrt{p}X, \sqrt{1-p}I\}^{\otimes 2(2k+1)}$.

We see that there are $4^{2k+1}$ combination of errors, for each of the cases, we have simulated the circuit and determined whether our code could correct the error or not. Fig. 9 illustrates our followed procedure in our simulation.

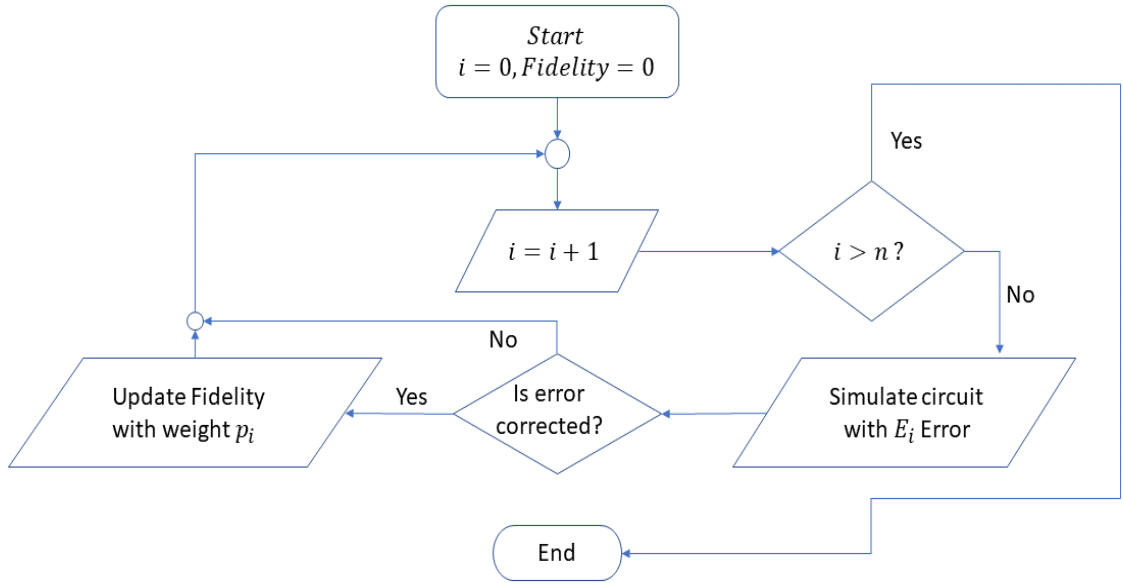

Fig. 9: Simulation Protocol for verification of theoretical results for fidelity. Channels are varied and different circuits are created for each channel. Finally, the expression for fidelity is generated computationally and matched with theoretical result.

Fig. 10 shows complete circuit for 5 qubit repetition code implemented to protect bell states:

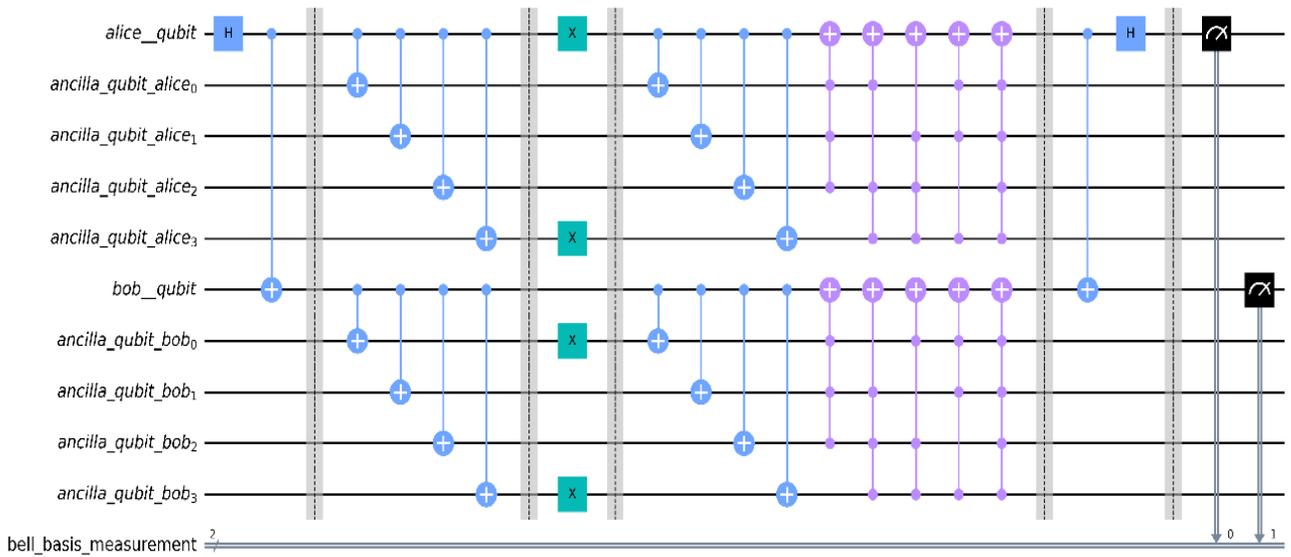

Fig. 10: Complete circuit implementing 5 qubit repetition code and bit flip channel for bell state.

Fig. 11 shows implementation of 3 qubit repetition code in bell state for phase flip protection. Similar procedures have been followed for bipartite product states.

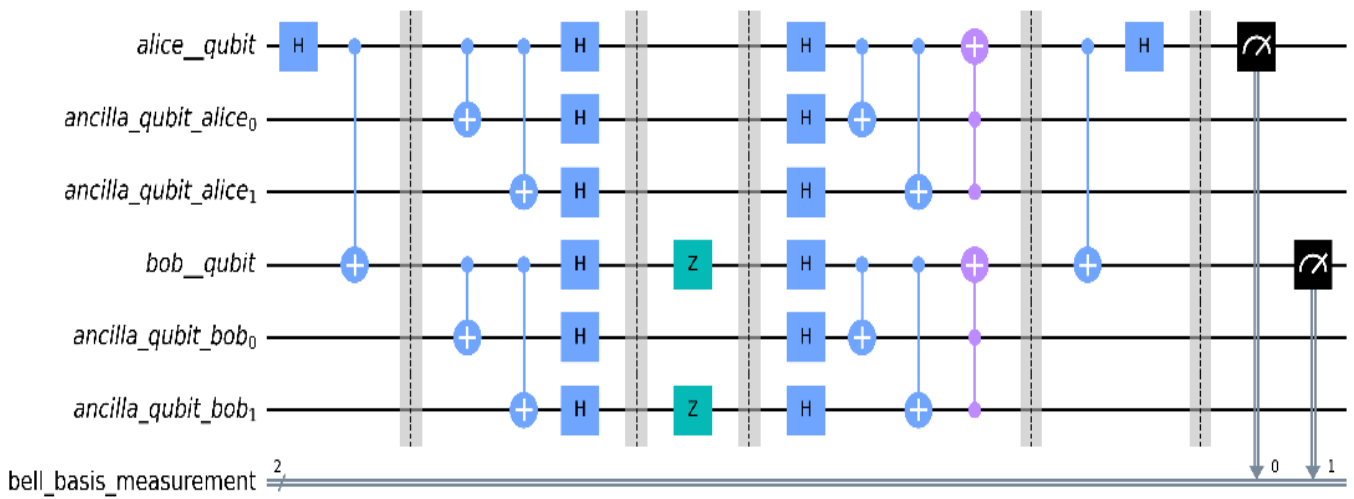

Fig. 11: Repetition Code for phase flip channel. Hadamard gate is applied before and after the channel.

We have theoretically analyzed the fidelity and numerically computed them using Qiskit QASM Simulator. Fig. 12 shows the result for 3 qubit repetition code implemented in both bell state and product state.

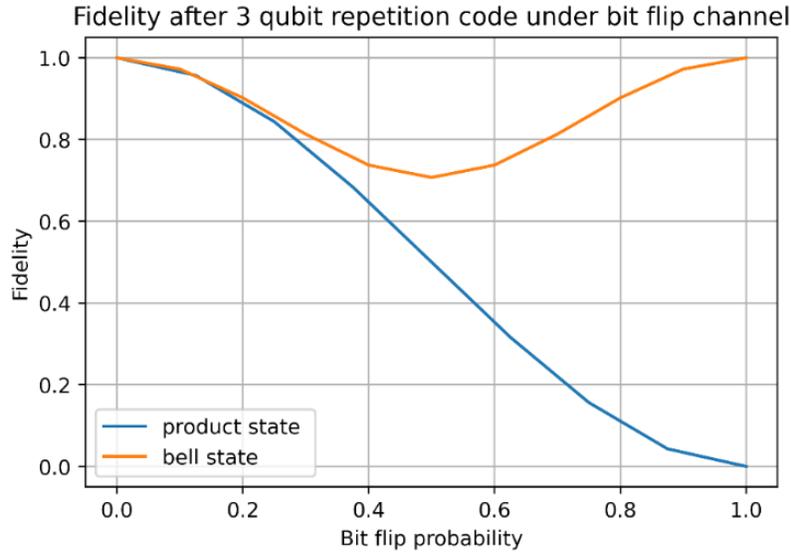

Fig. 12: Product state vs Bell State. We note that fidelity is better for bell states, and the curve is symmetric.

In our previous work, we implemented 3 and 5 qubit QRC for Bell States. The circuits were simulated using IBM QASM simulator, which is generalized analytically in this work, and our theoretical result perfectly agrees with our previous work result. In Fig. 13, we have shown the simulation result from our previous work [20].

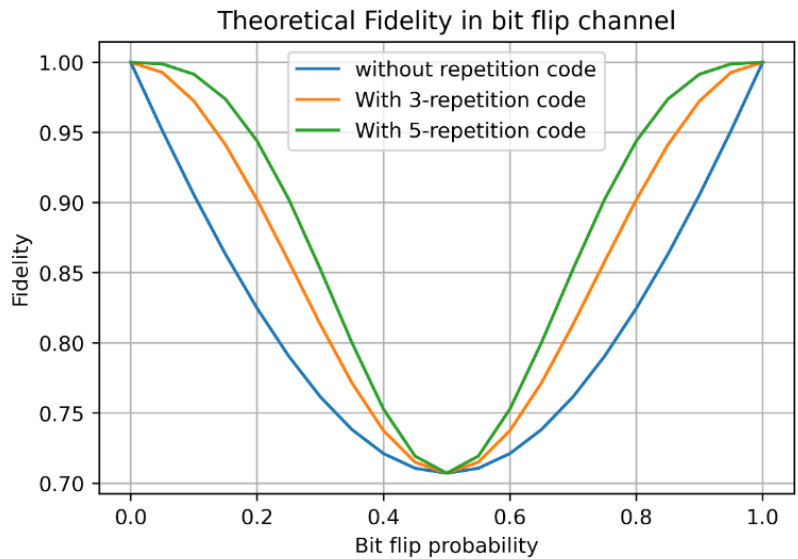

Fig. 13: Fidelity of Bell state using repetition code. We note that fidelity increases with the order.

Our calculated theoretical value of fidelity, $F_{bp}$, for implementing $(2k+1)$ repetition code in bipartite state is given below:

$$F_{bp} = \sqrt{\sum_{i=0}^{2(2k+1)} f(i) \times p^i (1-p)^{2(2k+1)-i}} \tag{23}$$

Here the coefficient $f(i)$ is given by:

$$f(i) = \begin{cases} \binom{2(2k+1)}{i}, & \text{for } i = 0,1,2,\ldots,k \\ \left(\sum_{j=1}^{i-1} \binom{2k+1}{j}\binom{2k+1}{i-j}\right), & \text{for } i = k+1,\ldots,2k \\ 0 \text{ for } i = 2k+1 \\ f(2(2k+1)-i), & \text{for } i > 2k+1 \text{ (bell state)} \\ 0 \text{ for } i > 2k+1 \text{ (product state)} \end{cases} \tag{24}$$

We see that the fidelity of bell state has a symmetric pattern, which we will discuss in the next section.

## 4. Fidelity Improvement for Short-distance Communication Using Stabilizer Formalism and A Classical Communication Based Protocol for Long-distance Communication

In this section, we shall decode the encoded noisy bell pairs using stabilizer formalism. We shall show that, in short-distance communication, we can correct all the bit flip errors but due to the requirement of nonlocal gate, we cannot do the same for long-distance communication. However, in the end of the section, we will propose a novel scheme where all the bit flip errors in long-distance communication too is correctable if a classical communication channel is available.

### 4.1 Stabilizer for QRC based Encoded Bell Pairs

We first note that, if a bell state is encoded using $(2k+1)$ repetition code, then $X^{\otimes 2(2k+1)}$ stabilizes the state. $|\phi_+\rangle = \langle X \otimes X, Z \otimes Z \rangle$ but $|\psi_+\rangle$ is stabilized by $Z \otimes Z$ with a global phase factor. Now, Eq. (11) states we can convert $|\phi_+\rangle$ state into other bell states and vice versa. Thus, we can convert other bell states into $|\phi_+\rangle$ state then encode before sending through channel.

Now, since $X^{\otimes 2(2k+1)}$ stabilizes encoded $|\phi_+\rangle$ state (let's denote by $|\phi_+^{2k+1}\rangle$), we will prove that two Hermitian and unitary Pauli error $E_1$ and $E_2$ will have the same effect if $E_1 E_2 = X^{\otimes 2(2k+1)}$.

Since $E_1 E_2 = X^{\otimes 2(2k+1)}$, we have $E_2 = E_1^\dagger X^{\otimes 2(2k+1)} = E_1 X^{\otimes 2(2k+1)}$

Now, $E_1 |\phi_+^{2k+1}\rangle = E_1 X^{\otimes 2(2k+1)} |\phi_+^{2k+1}\rangle$ or $E_1 |\phi_+^{2k+1}\rangle = E_2 |\phi_+^{2k+1}\rangle$

So, bit flip errors are tensor product of some $X$ and $I$, therefore they are both unitary and Hermitian (since $(A \otimes B)^\dagger = A^\dagger \otimes B^\dagger$). It explains the reason why in Eq. (24), $f(i) = f(2(2k+1) - i)$, $[i > 2k+1]$ for bell states although $f(i) = 0$, $[i > 2k+1]$ for arbitrary product states.

After some quick calculation we obtain a general representation of $|\phi_+^{2k+1}\rangle$ as,

$$|\phi_+^{2k+1}\rangle = \frac{1}{\sqrt{2}} \left( |0\rangle^{\otimes 2k+1} |0\rangle^{\otimes 2k+1} + |1\rangle^{\otimes 2k+1} |1\rangle^{\otimes 2k+1} \right) = \langle Z_1 Z_2, Z_1 Z_3, \ldots, Z_1 Z_{2(2k+1)}, X^{\otimes 2(2k+1)} \rangle$$

There can be separate representation of generators of stabilizer. For example, in 3 qubit repetition code, $|\phi_+^3\rangle = \langle Z_1 Z_2, Z_1 Z_3, Z_1 Z_4, Z_1 Z_5, Z_1 Z_6, X^{\otimes 6} \rangle = \langle Z_1 Z_2, Z_1 Z_3, Z_4 Z_5, Z_5 Z_6, Z_3 Z_4, X^{\otimes 6} \rangle$

But it is important to note that $|\phi_+^{2k+1}\rangle$ has total $2(2k+1)$ generators in stabilizer group. Therefore, it has the ability to detect $4^{(2k+1)}$ different type of errors. In addition, if two errors have the same error (i.e., product of these errors are in stabilizer group), then they have the same effect, which enables us to correct even more than $4^{(2k+1)}$ errors. Now, we note that for $(2k+1)$ repetition code implemented on bell states, there Krauss operator for bit flip channel is given by $E^b = \{\sqrt{p}X, \sqrt{1-p}I\}^{\otimes 2(2k+1)}$. The cardinality is given by $|E_b| = 2^{2(2k+1)} = 4^{(2k+1)}$. But we have proved that half of them can be written as $E_2 = E_1 X^{\otimes 2(2k+1)}$. Therefore, the number of unique bit flip errors are $4^{2k}$, and thus we have the ability to differentiate $4^{2k}$ more.

### 4.2 Computational Results for Stabilizer Formalism in Encoded Bell Pairs and Error Correcting Algorithm

We have derived the stabilizers for encoded Bell pairs in the previous section. In this section will show that this encoding can detect all possible bit flip errors separately. In this regard, we have encoded the Bell pair using 3 qubit repetition code.

After encoding, we have varied our channel (following our scheme in figure 8) to consider all possible bit flip errors and determine corresponding syndrome.

For syndrome measurement we have implemented the subcircuit of Fig. 5 for all the stabilizers: $Z_1Z_2$, $Z_1Z_3$, $Z_1Z_4$, $Z_1Z_5$, $Z_1Z_6$, $X^{\otimes 6}$. For this particular set of stabilizers, we have also provided an Error Correcting Algorithm as well. Fig. 14 shows the circuit implementation for encoded bell state with subspace measurement after channel for 3 qubit repetition code.

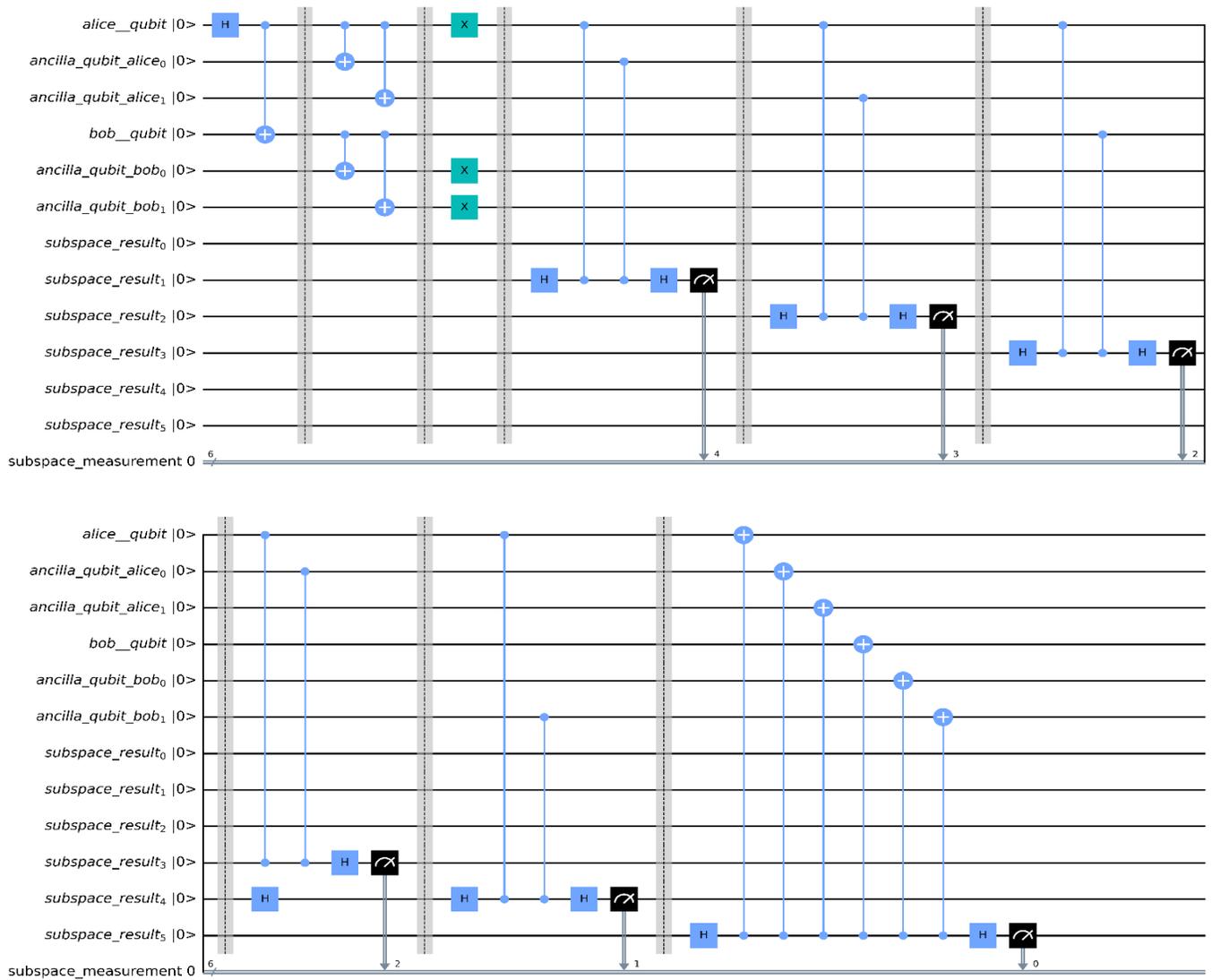

Fig. 14: Stabilizer syndrome detection for encoded bell state. Total 6 syndromes are measured here.

Table 1 lists all the bit flip errors with the syndromes.

**Table 1**

| Error --> Syndrome | Error --> Syndrome |
|---|---|
| $I\,I\,I\,I\,I\,I$ --> +1 +1 +1 +1 +1 +1 | $I\,I\,X\,X\,I\,I$ --> +1 −1 −1 +1 +1 +1 |
| $X\,I\,I\,I\,I\,I$ --> −1 −1 −1 −1 −1 +1 | $I\,I\,X\,I\,X\,I$ --> +1 −1 +1 −1 +1 +1 |
| $I\,X\,I\,I\,I\,I$ --> −1 +1 +1 +1 +1 +1 | $I\,I\,X\,I\,I\,X$ --> +1 −1 +1 +1 −1 +1 |
| $I\,I\,X\,I\,I\,I$ --> +1 −1 +1 +1 +1 +1 | $I\,I\,I\,X\,X\,I$ --> +1 +1 −1 −1 +1 +1 |
| $I\,I\,I\,X\,I\,I$ --> +1 +1 −1 +1 +1 +1 | $I\,I\,I\,X\,I\,X$ --> +1 +1 −1 +1 −1 +1 |
| $I\,I\,I\,I\,X\,I$ --> +1 +1 +1 −1 +1 +1 | $I\,I\,I\,I\,X\,X$ --> +1 +1 +1 −1 −1 +1 |
| $I\,I\,I\,I\,I\,X$ --> +1 +1 +1 +1 −1 +1 | $X\,X\,X\,I\,I\,I$ --> +1 +1 −1 −1 −1 +1 |
| $X\,X\,I\,I\,I\,I$ --> +1 −1 −1 −1 −1 +1 | $X\,X\,I\,X\,I\,I$ --> +1 −1 +1 −1 −1 +1 |
| $X\,I\,X\,I\,I\,I$ --> −1 +1 −1 −1 −1 +1 | $X\,X\,I\,I\,X\,I$ --> +1 −1 −1 +1 −1 +1 |
| $X\,I\,I\,X\,I\,I$ --> −1 −1 +1 −1 −1 +1 | $X\,X\,I\,I\,I\,X$ --> +1 −1 −1 −1 +1 +1 |
| $X\,I\,I\,I\,X\,I$ --> −1 −1 −1 +1 −1 +1 | $X\,I\,X\,X\,I\,I$ --> −1 +1 +1 −1 −1 +1 |
| $X\,I\,I\,I\,I\,X$ --> −1 −1 −1 −1 +1 +1 | $X\,I\,X\,I\,X\,I$ --> −1 +1 −1 +1 −1 +1 |
| $I\,X\,X\,I\,I\,I$ --> −1 −1 +1 +1 +1 +1 | $X\,I\,X\,I\,I\,X$ --> −1 +1 −1 −1 +1 +1 |
| $I\,X\,I\,X\,I\,I$ --> −1 +1 −1 +1 +1 +1 | $X\,I\,I\,X\,X\,I$ --> −1 −1 +1 +1 −1 +1 |
| $I\,X\,I\,I\,X\,I$ --> −1 +1 +1 −1 +1 +1 | $X\,I\,I\,X\,I\,X$ --> −1 −1 +1 −1 +1 +1 |
| $I\,X\,I\,I\,I\,X$ --> −1 +1 +1 +1 −1 +1 | $X\,I\,I\,I\,X\,X$ --> −1 −1 −1 +1 +1 +1 |

We note that each of the errors has a distinct set of stabilizers. So, we should be able to correct them also. Keeping in mind that $X_i$ error ($X_i = I \otimes ... \otimes X \otimes ... \otimes I$, X in the $i^{th}$ position and rest is I) and $X^{\otimes 2(2k+1)} X_i = X \otimes ... \otimes I \otimes ... X$ are same error, we note that, if we rotate the syndromes clockwise, then we put classically control NOT gate in the position where $c_i = +1$, then the error becomes canceled.

For example, if the syndrome is $\{+1\ -1\ +1\ -1\ -1\ +1\}$ then we first rotate clockwise giving us $\{+1 +1\ -1\ +1\ -1\ -1\}$. Now +1 is in the position 1,2,4. From table 2, we see this is the syndrome for the error $X_1 X_2 X_4 = XXIXII$. Since the errors are both unitary and Hermitian, we can apply classically controlled NOT gate in position 1,2,4 to nullify the error. Note that according to Eq. (42), $f(3) = 0$ for 3 qubit repetition code. But using the stabilizer formalism, we have been able to distinguish the error as well as correct it. Thus, Bell states are in fact able to correct all possible bit flip errors.

Another example is, for syndrome $\{+1\ +1\ +1\ -1\ +1\ +1\}$, by rotating we get $\{+1 +1\ +1 + 1\ -1\ +1\}$, which represents $X_1 X_2 X_3 X_4 X_6$ error, hence also $X_5$ errors.

## 4.3 Encoded Bell in Short and Long-Distance Communication: A Comparison

In section 4.2, we have demonstrated that every possible bit-flip error is correctable. However, it doesn't mean we can always measure all the syndromes, which is a requirement for 100% fidelity. For short-distance, the local encoded bell pairs can provide all the syndrome measurement. But for long-distances, it is not the case.

At first, we will analyze some points:

1) For product states, the stabilizer set can be written in the following form:
$\langle Z_1 Z_2, Z_1 Z_3, \ldots, Z_1 Z_{2k+1}, Z_{2k+2} Z_{2k+3}, Z_{2k+2} Z_{2k+4}, \ldots, Z_{2k+2} Z_{2(2k+2)} \rangle$
The cardinality is $16^k$. Therefore for 3 qubit repetition code, there can be 16 distinct syndromes only. And we already know that it can correct exactly $(6C_0 + 6C_1 + 3C_1 \times 3C_1) = 16$ errors. Thus, the decoder for repetition code is sufficient for the product state. But for bell states, the decoder circuit is not sufficient at all, using stabilizer formalism we can, in fact, correct all possible bit flip errors.

2) For phase flip errors, even the number of phase flips are already in stabilizer set of encoded bell state, hence correcting these errors are not necessary, in fact they don't result in errors at all. Therefore, we can eventually distinguish $2^{2k-1}$ more error which are not only bit flip or phase flip. It gives us a huge list of error correcting ability.

3) For encoded Bell pairs, the stabilizer generators are given by the following:
$\langle Z_1 Z_2, Z_1 Z_3, \ldots, Z_1 Z_{2k+1}, Z_{2k+2} Z_{2k+3}, Z_{2k+2} Z_{2k+4}, \ldots, Z_{2k+2} Z_{2(2k+2)}, Z_{2k+1} Z_{2k+2}, X^{\otimes 2(2k+1)} \rangle$
They are very similar to product state except there are two extra stabilizers: $Z_{2k+1} Z_{2k+2}, X^{\otimes 2(2k+1)}$. Due to $X^{\otimes 2(2k+1)}$, we get a symmetric curve for fidelity. But for generic bipartite entangled state (not Bell state), it is not a stabilizer. But we still have an extra stabilizer: $Z_{2k+1} Z_{2k+2}$, which required control gate between both Alice and Bob's side, which is indeed a non-local operation. Therefore, for long-distance communication, we cannot correct some errors due to our inability to perform such operation.

From cases (1) and (3), we understand that entanglement state in general behaves same as product states for long-distance communication. Therefore, fidelity is also limited by repetition code fidelity. However, for local entangled states, we can correct more errors.

## 4.4 Encoded Bell in Long-Distance Communication with Classical Communication Channel

Suppose, Alice and Bob share an encoded $|\phi_+^3\rangle$ state.

$$|\phi_+^3\rangle = \langle Z_1 Z_2, Z_1 Z_3, Z_4 Z_5, Z_5 Z_6, Z_3 Z_4, X^{\otimes 6} \rangle$$

We know, all the bit flip errors commute with $X^{\otimes 6}$, but to implement the stabilizer error correction, we need to know the syndrome of $Z_3 Z_4$, but Alice and Bob live far away from each other. Alice can measure the subspace of $Z_1 Z_2, Z_1 Z_3$ and Bob can measure $Z_4 Z_5, Z_5 Z_6$ by using local operation. However, due to the non-locality of control $Z_3 Z_4$, we can't correct half of the bit flip errors, thus giving a lower fidelity. Only due to $X^{\otimes 6}$ stabilizer, the decoder of 3 qubit repetition code can correct twice error than product states.

However, we note that, if $EZ_3 = m_1 Z_3 E$ and $EZ_4 = m_2 Z_4 E$, $(m_1, m_2 \in \{-1, +1\})$ then

$$E(Z_3 Z_4) = (EZ_3)Z_4 = m_1 Z_3(EZ_4) = m_1 m_2 (Z_3 Z_4) E$$

(Since Pauli group elements either commute or anti-commute.)

So, if Alice could measure whether $E$ commutes with $Z_3$ and Bob could measure whether $E$ commutes with $Z_4$, then by using classical communication, they should be able to determine whether $E$ commutes with $Z_3 Z_4$. We note that $E, Z_3$ and $Z_4$ are Hermitian. So, we define two set of observables given by anti-commutators:

$$O_1 = \{E, Z_3\} = (1 + m_1) Z_3 E, O_2 = \{E, Z_4\} = (1 + m_2) Z_4 E$$

It is easy to show that both $O_1$ and $O_2$ are Hermitian. Now anti-commutator is 0, when two Hermitian anti-commutes with each other. Keeping that in mind let us calculate $O = [O_1, O_2]$. If $O = 0$ then there exist a simultaneous eigen-basis where we can measure both $O_1, O_2$ simultaneously. Now,

$$O = [O_1, O_2] = (1 + m_1)(1 + m_2)(m_2 - m_1) Z_3 Z_4 \tag{25}$$

We note that for $(m_1, m_2) \in \{(-1, -1), (-1, +1), (+1, -1), (+1, +1)\}$, Eq. (25) always results in 0. Therefore, we can always measure both $O_1$, $O_2$ simultaneously, thus measure $m_1 m_2$ by using classical communication. This enables us to obtain all the necessary stabilizer syndrome.

So, if we want to obtain similar fidelity in long-distance communication, then we can obtain this using classical communication. Figure 3 states our proposed protocol for long-distance quantum communication with bit flip or phase flip quantum channel along with a classical communication channel.

For arbitrary quantum channel, we need to encrypt Bell pair using nested or other form of encoding process. After that the similar procedure can be followed.

## 5. Quantum Thermodynamical Reasoning Behind Fidelity Improvement

Von Neumann Entropy is defined as $S(\rho) = -Tr[\rho \log_2 \rho]$. For pure state the entropy is zero. So, for both bipartite product state and bipartite entangled state of particle A, B, we have $S_{AB} = 0$.

Now we define conditional entropy, $S_{A|B} = S_{AB} - S_A$. If we assume A particle is Alice's and B is Bob's, then $S_{A|B}$ is the information that Bob doesn't know after observing his qubit.

Similarly, we define mutual information as, $I(A, B) = S_A - S_{A|B}$, which is the information we can learn about A by observing B, and vice versa [Since $I(A, B) = I(B, A)$].

For product state, $\rho_{AB} = \rho_A \otimes \rho_B$ and $\rho_A, \rho_B$ are pure states. Hence $S_A = S_B = 0$ too. Therefore, both conditional and mutual information is zero, i.e., $S_{A|B} = 0, I(A, B) = 0$.

However, for entangled state we have $S_{A|B} = -S_A < 0$ and $I(A, B) = S_A + S_B > 0$

Therefore, for entangled states we have some mutual information. When we designed the QRC, we created entanglement between the main qubit and ancilla qubits. Due to the mutual information created in this process, we were able to reduce the error after passing through the channel. When the main qubit gets entangled with ancilla, the Von Neumann Entropy is zero. However, after passing through channel, the state becomes a mixed state and the Von Neumann Entropy rises. But during the decoding, Mutual Information helps to get rid of the errors, therefore mutual information is a key point for Quantum Error Correction.

However, to retrieve the benefit of it, we demand the whole system to be localized, not separated by long-distance. This is the reason why entangled systems separated by long-distance doesn't help much, the mutual information cannot be retrieved without any non-local operation. The more mutual information the system has, the more errors can be corrected using proper encoding. In stabilizer formalism, this was related to the number of stabilizer generators, which depends on encoding as well as the mutual information.

6. **Conclusion**

In this work, the error correction in entangled state is studied in detail from different angles and the proposed results suggest the significant improvement of fidelity for all the cases. We have investigated how the error correcting ability changes in different scenarios: short-distance communication, long-distance communication (no classical communication), and long-distance communication with classical communication channel. The idea of long and short distance communication is used from laboratory perspective. If the entangled state is local or can be localized at distance so that our equipment's are capable of performing controlled unitary operation, then we have considered it short distance. When we cannot bring the entangled states closer to perform control unitary measurements, then we call them to be located at long distance. The error correcting abilities seem to change significantly for long and short-distance. As a result, separate models subjected to QRC based encoding has been proposed for long and short-distance. Using Stabilizer Formalism, we have demonstrated that in long-distance communication, fidelity is limited by QRC, but the fidelity improvement after channel noise is significant. A novel model using local measurement and classical communication for long-distance communication has been proposed with proper analytical proof. We have shown how encoded local entanglement state can correct all single bit flip or phase flip errors, which makes an enormous improvement in fidelity compared to QRC. This result is very strong and shows how powerful Error Correcting codes are and how entanglement actually increases the error correcting ability. We have generalized all the circuits and protocols; then calculated the fidelity for both single qubit and bipartite quantum state. We have verified our analytical results using QASM Simulator. The described procedure can be quite impactful for QECC based applications and computations. In further endeavor, we would like to extend our work for arbitrary quantum channel. Results, presented in this work, can be imperative for designing new generation quantum repeaters and other relevant applications.

**Acknowledgement**

The authors would like to thank Mr. Sowmitra Das, Lecturer, Department. of CSE, BRAC University for his valuable suggestions and constant support. M.R.C. Mahdy acknowledges the support of NSU CTRGC grant 2022-23 (approved by the members of BOT) and the internal grant of North South University.

<div align="center">**References**</div>